\documentclass{Interspeech}
\usepackage{svg,multirow,adjustbox,threeparttable,amsmath, amssymb,pifont,arydshln,url}

\interspeechcameraready

\title{SuPseudo: A Pseudo-supervised Learning Method for Neural Speech Enhancement in Far-field Speech Recognition}

\author[affiliation={1}]{Longjie}{Luo}
\author[affiliation={1*}]{Lin}{Li}
\author[affiliation={2*}]{Qingyang}{Hong}

\affiliation{School of Electronic Science and Engineering}{Xiamen University}{China}
\affiliation{School of Informatics}{Xiamen University}{China}
\email{\{lilin,qyhong\}@xmu.edu.cn}

\usepackage{comment}

\begin{document}

\maketitle

\makeatletter
\def\@makefnmark{}
\footnotetext{* Corresponding author.}
\footnotetext{This work was supported in part by the National Natural Science Foundation of China under Grants 62371407 and 62276220, and the Innovation of Policing Science and Technology, Fujian province (Grant number: 2024Y0068).}
\def\@makefnmark{\hbox{\@textsuperscript{\normalfont\@thefnmark}}} % 恢复默认脚注样式
\makeatother

% the abstract here must exactly match the abstract entered into the paper submission system
%  which consists of \textbf{T}ime alignment, \textbf{L}evel alignment, and \textbf{S}ignal-to-noise ratio filtering modules,
\begin{abstract}
Due to the lack of target speech annotations in real-recorded far-field conversational datasets, speech enhancement (SE) models are typically trained on simulated data. However, the trained models often perform poorly in real-world conditions, hindering their application in far-field speech recognition. To address the issue, we (a) propose \textit{direct sound estimation} (DSE) to estimate the oracle direct sound of real-recorded data for SE; and (b) present a novel pseudo-supervised learning method, SuPseudo, which leverages DSE-estimates as pseudo-labels and enables SE models to directly learn from and adapt to real-recorded data, thereby improving their generalization capability. Furthermore, an SE model called FARNET is designed to fully utilize SuPseudo. Experiments on the MISP2023 corpus demonstrate the effectiveness of SuPseudo, and our system significantly outperforms the previous state-of-the-art. A demo of our method can be found at \url{https://EeLLJ.github.io/SuPseudo/}.
\end{abstract}

\section{Introduction}
Speech enhancement (SE) algorithms can significantly improve automatic speech recognition (ASR) performance in far-field conversational scenarios \cite{boeddeker2018front}. Meanwhile, modern data-driven SE models have showcased powerful performance \cite{hao2021fullsubnet, yu2023efficient, abdulatif2024cmgan}, attracting increasing researchers to explore their application in real-world far-field speech recognition \cite{wu2024multimodal,vinnikov2024notsofar}. Since existing real-recorded far-field conversational datasets, such as CHiME-5/-6/-7 \cite{barker2018fifth,watanabe2020chime,cornell2023chime} and MISP2021/2022/2023 \cite{chen2022first,wang2023multimodal,wu2024multimodal}, do not provide any direct annotations for SE, the models are typically trained on simulated noisy-clean data pairs. However, due to Lombard effect \cite{michelsanti2019effects}, microphone clipping effect, room geometry, etc, there are significant mismatches between the acoustic properties of simulated and real-recorded data (i.e., domain mismatch), leading to severe performance degradation of the trained SE models in real-world conditions \cite{leglaive2023chime}. Worse still, compared with human ears, ASR back-ends are more sensitive to this performance drop. Hence, leveraging modern data-driven SE models as front-ends for real-world far-field speech recognition still faces challenges \cite{barker2018fifth,watanabe2020chime,cornell2023chime,chen2022first,wang2023multimodal}.

To mitigate the domain mismatch issue, in \cite{wu2024multimodal,han2024audio}, SE models are jointly optimized with ASR back-ends on real-recorded data by leveraging the weak supervision of word transcriptions. Additionally, the perceptual evaluation of speech quality (PESQ) estimators \cite{xu2024employing} and the DNSMOS models \cite{reddy2022dnsmos} can also provide reference-free losses for fine-tuning SE models on real-recorded data. Recently, inspired by the blind deconvolution problem, Wang \cite{wang2024superm2m} designed some loss functions that enforce the model-estimates to be reconstructed to real-recorded data, effectively utilizing real-recorded data in the single-speaker condition.

\begin{figure}
    \centerline{\includegraphics[width=0.75\columnwidth]{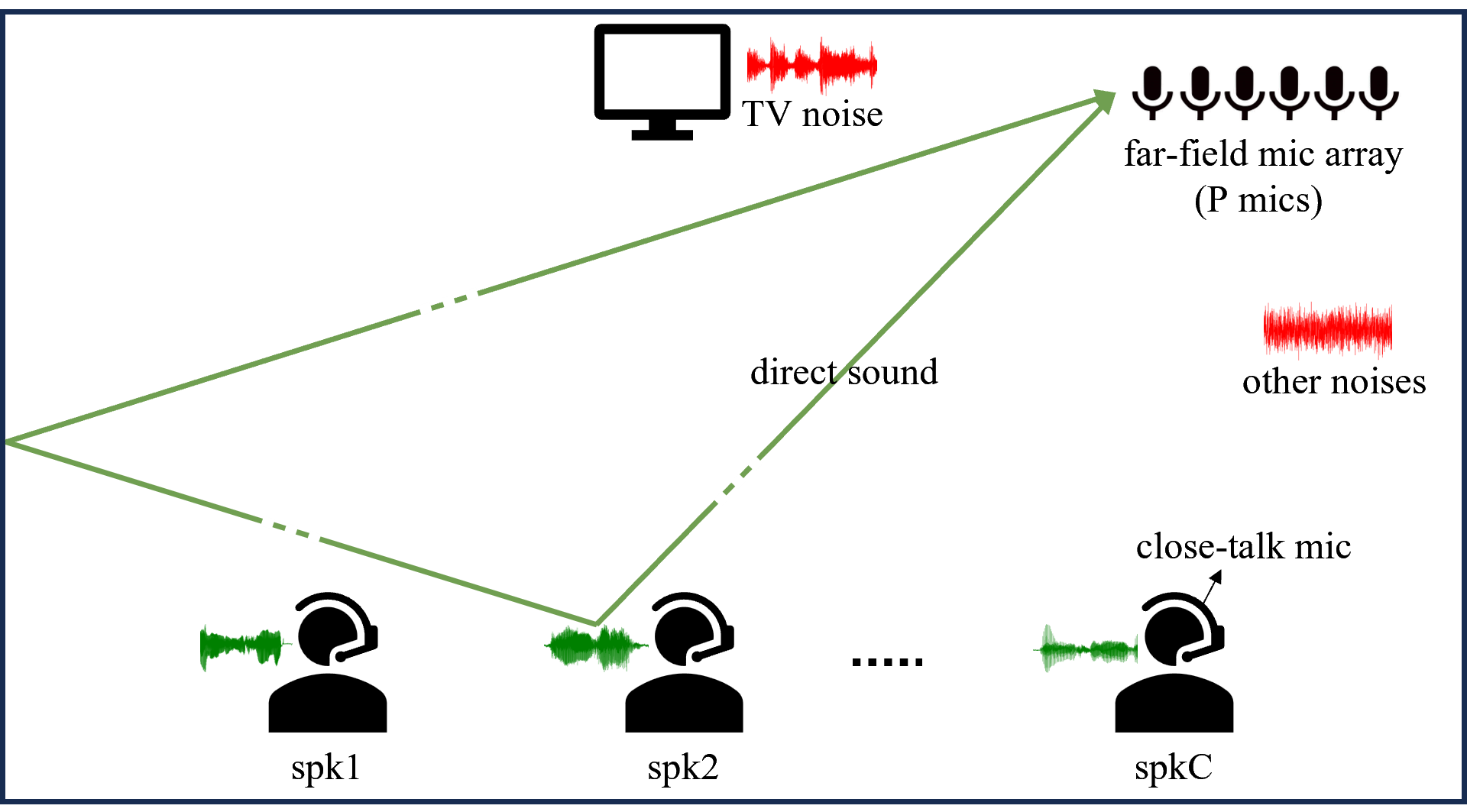}}
    \caption{Schematic diagram of the recording scenario.}
    \label{fig1}
    \vspace{-15pt} % 减少图片下方的间距
\end{figure}

Unlike previous works, this paper tries directly estimating the oracle direct sound (as illustrated in Fig. \ref{fig1}) of real-recorded far-field conversational datasets \cite{chen2022first,wang2023multimodal,wu2024multimodal} for SE. Since close-talk recordings, which are used for manual transcription, typically have very high signal-to-noise ratios (SNRs), it is reasonable to use them for estimating the oracle direct sound. However, (a) close-talk recordings may still contain a small amount of speech leakage; and (b) close-talk recordings are time- and level-misaligned with the oracle direct sound due to signal delay and attenuation from long-distance propagation. To address these issues, we first mask the close-talk recordings using timestamps, and then we present an effective method called \textit{direct sound estimation} (DSE), which performs multi-frame linear filtering on close-talk speech, to achieve alignment.

\begin{figure}
    \centerline{\includegraphics[width=\columnwidth]{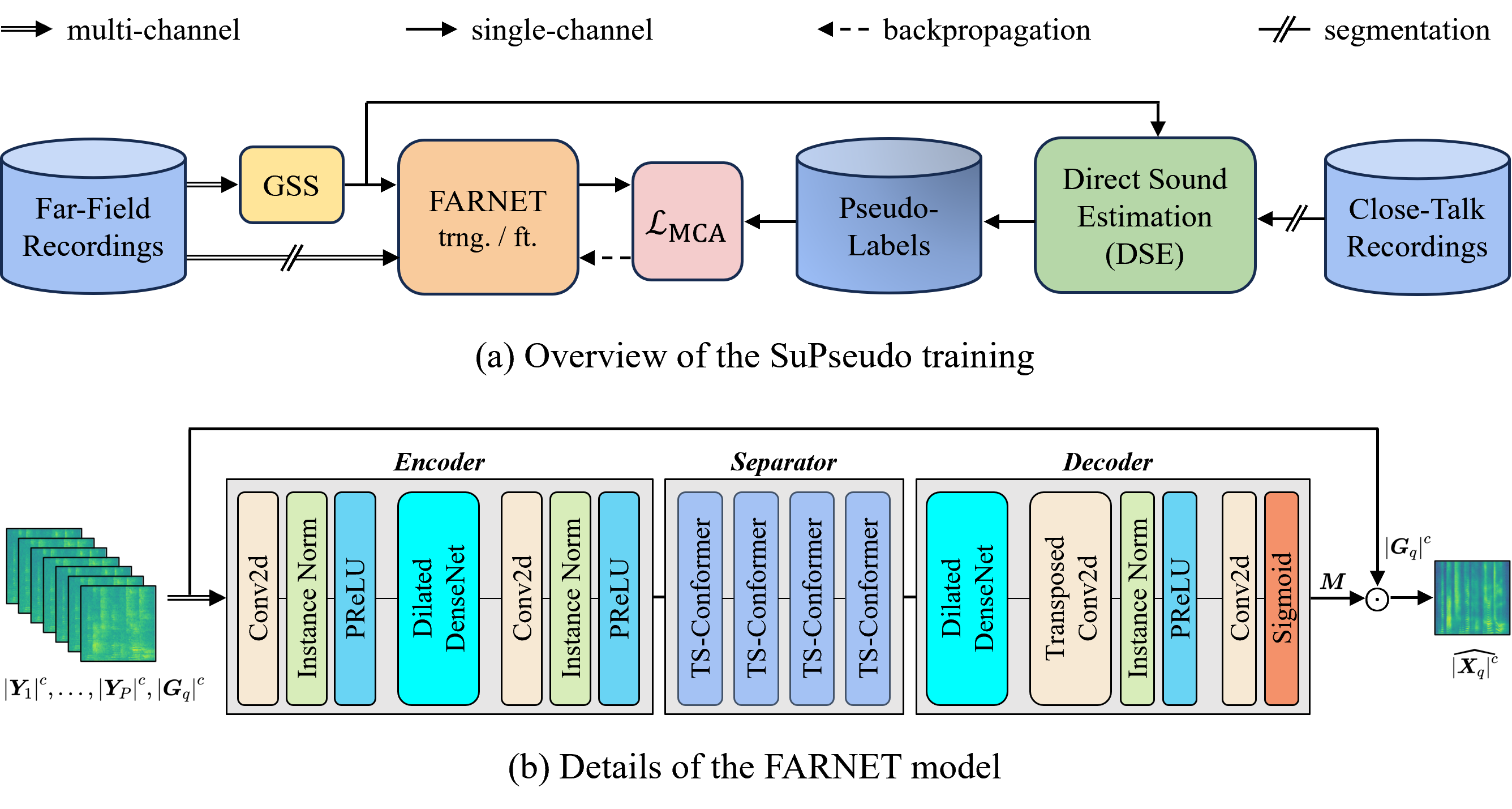}}
    \caption{Illustration of the proposed method. In SuPseudo, if the model has been pre-trained on simulated data, then the fine-tuning (ft.) is performed; otherwise, the training (trng.)  directly on real-recorded data is performed.}
    \label{fig2}
    \vspace{-15pt}
\end{figure}

Based on the estimated direct sound of real-recorded data, we then design a novel training framework called SuPseudo (as shown in Fig. \ref{fig2}(a)). In SuPseudo, the direct-sound-estimates (DSE-estimates) are treated as pseudo-labels for training (referred to as \textit{pseudo-supervised learning} \cite{lv2021pseudo}), enabling SE models to directly learn from and adapt to real-recorded data. To against the potential errors in pseudo-labels during training, we design a new loss function called \textit{magnitude constraint adjustable} (MCA) loss. Moreover, by incorporating a supervised pre-training stage using massive simulated data, the performance of SuPseudo is further improved. Although SuPseudo is applicable to all SE models, to better match far-field speech recognition \cite{barker2018fifth,watanabe2020chime,cornell2023chime,chen2022first,wang2023multimodal,wu2024multimodal}, we design a new SE model called FARNET (as shown in Fig. \ref{fig2}(b)). We validate our method on MISP2023 corpus \cite{wu2024multimodal}, a real-recorded far-field multi-speaker conversational corpus, and our system greatly outperforms the previous state-of-the-art (SOTA). The main contributions of this paper are as follow:
\begin{itemize}
	\item To our knowledge, we are the first to present an effective method, DSE, for estimating the oracle direct sound of real-recorded data.
    \item We propose SuPseudo to utilize DSE-estimates and enable SE models to directly learn from real-recorded data.
    \item The proposed FARNET system trained using SuPseudo greatly surpasses previous SOTA on MISP2023 Challenge.
\end{itemize}

\section{Problem Formulation}
\label{sec2}
As shown in Fig. \ref{fig1}, $N$ speakers are spontaneously grouped for daily conversation in a reverberant room. Each speaker wears a high-fidelity microphone (i.e., close-talk microphone) positioned at the center of their chin. Additionally, a $P$-channel microphone array (i.e., far-field microphone array) is placed near a TV. Other background noises, such as air conditioner and movement noise, are also present in the room.

The signal collected by a far-field microphone $p \in \left \{ 1,...,P \right \}$ can be formulated in the short-time Fourier transform (STFT) domain as
\small % 这里设置整个公式的字体大小
\begin{equation}
\begin{split}
Y_{p}(t, f) & = \sum_{n=1}^{N} R_{n, p}(t, f)+V_{p}(t, f),
\end{split}
\end{equation}
\normalsize % 恢复正常字体大小
where $ R_{n, p}(t, f) $ and $ V_{p}(t, f) $ respectively denote the STFT coefficients of the reverberant speech of speaker $n\in\left \{ 1,...,N \right \}$ and the noise at time frame $t$ and frequency subband $f$. 

Then we use guided source separation (GSS) \cite{raj23_interspeech}, along with the oracle diarization, to pre-process the raw $P$-channel far-field recordings $\left \{ Y_{1}(t, f),...,Y_{P}(t, f) \right \}$, initially separating out the speech segments for each speaker. The GSS-separated speech segments of target speaker $q$ (each speaker is sequentially treated as a target speaker) can be formulated as
\small % 这里设置整个公式的字体大小
\begin{equation}
\begin{split}
G_{q}(t, f) & = R_{q}(t, f)+I_{q}(t, f) \\ 
& =\boldsymbol{h}_{q}^{d}(f)^{\mathrm{H}} \bar{\boldsymbol{S}}_{q}(t, f)+\boldsymbol{h}_{q}^{e,l}(f)^{\mathrm{H}} \widetilde{\boldsymbol{S}}_{q}(t, f)+I_{q}(t, f), 
\end{split}
\label{eq:GSS_signal}
\end{equation}
\normalsize % 恢复正常字体大小
where $R_{q}(t, f)$ and $I_{q}(t, f)$ denote the STFT coefficients of reverberant target speech and residual interferences (i.e., reverberant interfering speech plus background noise), respectively. $\boldsymbol{h}_{q}^{d}(f)^{\mathrm{H}} \bar{\boldsymbol{S}}_{q}(t, f)$ denotes the narrowband approximation \cite{talmon2009relative} of target direct sound, with $\boldsymbol{h}_{q}^{d}(f) \in \mathbb{C}^{L}$ being a linear filter and $\bar{\boldsymbol{S}}_{q}(t, f) = \left [ S_{q}(t,f),...,S_{q}(t-L+1,f) \right ] ^{T}$ stacking $L$ delayed copies of the dry source signal $S_{q}(t,f)$. Similarly, $\boldsymbol{h}_{q}^{e,l}(f)^{\mathrm{H}} \widetilde{\boldsymbol{S}}_{q}(t, f)$ denotes the narrowband approximation of target non-direct sound (i.e., early reflections plus late reverberation), with $\boldsymbol{h}_{q}^{e,l}(f) \in \mathbb{C}^{K}$ ($K> L$) being a linear filter and $\widetilde{\boldsymbol{S}}_{q}(t, f) = \left [ S_{q}(t,f),...,S_{q}(t-K+1,f) \right ] ^{T}$ stacking $K$ delayed copies of the dry source signal $S_{q}(t,f)$.

Additionally, the corresponding close-talk speech segments, obtained by segmenting the close-talk recordings of the target speaker $q$ using timestamps, can be formulated as
\small % 这里设置整个公式的字体大小
\begin{equation}
\begin{split}
Y_{q}(t, f) & \approx S_{q}(t, f).
\end{split}
\label{eq:near_signal}
\end{equation}
\normalsize % 恢复正常字体大小
Since the segmented speech is nearly clean (rarely interfered by the off-target sources, as noted in \cite{wu2024multimodal}) and the close-talk microphone is close enough to the speech source, $Y_{q}(t, f)$ is considered the dry source signal $S_{q}(t, f)$ in this paper.

This paper aims to further reconstruct the target speaker $q$'s direct sound $\boldsymbol{h}_{q}^{d}(f)^{\mathrm{H}} \bar{\boldsymbol{S}}_{q}(t, f)$ from the GSS-output $G_{q}(t, f)$, so as to improve the speech quality and the ASR performance.

\section{Proposed Method}
Fig. \ref{fig2} illustrates the proposed SuPseudo training and FARNET model. As described in Section \ref{sec2}, we first use GSS \cite{raj23_interspeech}, along with the oracle diarization, to pre-process $P$-channel far-field recordings. The proposed FARNET model is then used to further enhance the GSS-output $G_{q}(t, f)$. For SuPseudo training, the DSE-estimates are used as pseudo-labels, and the model is trained directly on real-recorded data by optimizing the MCA loss. As an option, two-stage SuPseudo can be used. In stage 1, the synthetic $P$-channel far-field mixtures and direct sound data pairs are used to optimize the mean square error (MSE) loss for supervised pre-training. In stage 2, the model is further fine-tuned on real-recorded data with DSE-estimates.

\subsection{FARNET}
Considering that ASR back-ends only utilize magnitude information, and based on our observation that the magnitude-masking method has greater robustness in challenging far-field acoustic environments, the proposed FARNET model adopts the magnitude-masking strategy, leaving the phase unaltered. 

Fig. \ref{fig2}(b) gives the details of the proposed FARNET model, which adopts an encoder-separator-decoder structure. Let $\left | \boldsymbol{Y}_{p}  \right | ^{c} \in \mathbb{R} ^{T\times F}$ (where $p \in \left \{ 1,...,P \right \}$) and $\left | \boldsymbol{G}_{q}  \right | ^{c} \in \mathbb{R} ^{T\times F} $, with $c=0.3$ following \cite{braun2021consolidated}, respectively denote the power-law compressed magnitude spectrogram of segmented $Y_{p}(t, f)$ and $G_{q}(t, f)$. Following \cite{luo2024xmuspeech}, we stack $\left | \boldsymbol{Y}_{1}  \right | ^{c},...,\left | \boldsymbol{Y}_{P}  \right | ^{c},\left | \boldsymbol{G}_{q}  \right | ^{c}$ along the channel dimension to obtain the model input $\boldsymbol{Y}_{in} \in \mathbb{R} ^{T\times F\times \left ( P+  1 \right ) }$, which is similar to ensemble learning techniques \cite{zhang2016deep}. The encoder consists of a dilated DenseNet \cite{pandey2020densely} and two convolution blocks, each comprising a convolution layer, an instance normalization \cite{ulyanov2016instance}, and a PReLU activation function. The encoder progressively maps the low-dimensional input to a high-dimensional feature space, where the non-target signal tends to be more separately distinguishable from the mixture $G_{q}(t, f)$. The separator employs four TS-Conformer blocks \cite{abdulatif2024cmgan}, which model global dependencies of high-dimensional features along both time and frequency dimensions to effectively estimate non-target signal. The decoder, employing modules similar to the encoder, reconstructs the corresponding magnitude mask $\boldsymbol{M} \in \mathbb{R} ^{T\times F}$ based on the non-target signal estimated by the separator. Finally, the enhanced power-law compressed magnitude spectrogram is obtained via a Hadamard product: $\widehat{\left | \boldsymbol{X}_{q}  \right | ^{c}} =\boldsymbol{M}\odot  \left | \boldsymbol{G}_{q}  \right | ^{c}$.

\subsection{DSE}
To address the performance degradation issue caused by the mismatches between simulated and real-recorded data (domain mismatch), here we directly estimate the oracle direct sound of real-recorded data (i.e., $\boldsymbol{h}_{q}^{d}(f)^{\mathrm{H}} \bar{\boldsymbol{S}}_{q}(t, f)$ in $G_{q}(t, f)$) for SE. Inspired by the forward convolutive prediction (FCP) \cite{wang2021convolutive}, we perform multi-frame linear filtering on close-talk speech segments $Y_{q}(t, f)$ to estimate the oracle direct sound $\boldsymbol{h}_{q}^{d}(f)^{\mathrm{H}} \bar{\boldsymbol{S}}_{q}(t, f)$. The filter is obtained by solving the following optimization problem:
\small % 这里设置整个公式的字体大小
\begin{equation}
\begin{split}
\hat{\boldsymbol{h}}_{q}^{d}(f)=\underset{\acute{\boldsymbol{h}}_{q}^{d}(f)}{\operatorname{argmin}} \sum_{t} \frac{\left|G_{q}(t, f)-\acute{\boldsymbol{h}}_{q}^{d}(f)^{\mathrm{H}} \breve{\boldsymbol{Y}}_{q}(t, f)\right|^{2}}{\lambda(t, f)},
\end{split}
\label{eq:filter_est}
\end{equation}
\normalsize % 恢复正常字体大小
where $\breve{\boldsymbol{Y}}_{q}(t, f)=[Y_{q}(t, f), \ldots, Y_{q}(t-\hat{L}+1, f)]^{\mathrm{T}} \in \mathbb{C}^{\hat{L}}$ stacks $\hat{L}$ T-F units, and $\hat{L}$, given by Eq. (\ref{eq:num_taps}), denotes the predicted time frames delay ($L$ is the oracle one) from the target speaker's dry source signal $S_{q}(t,f)$ to direct sound signal $\boldsymbol{h}_{q}^{d}(f)^{\mathrm{H}} \bar{\boldsymbol{S}}_{q}(t, f)$. $\lambda(t, f)$, given by Eq. (\ref{eq:lambda}), is a weighting term.

The predicted time frames delay, which is also the order of filter $\acute{\boldsymbol{h}}_{q}^{d}(f)$, is calculated as
\small % 这里设置整个公式的字体大小
\begin{equation}
\begin{split}
\hat{L}=\left \lceil \frac{D}{a \times H} \right \rceil +1,
\end{split}
\label{eq:num_taps}
\end{equation}
\normalsize % 恢复正常字体大小
where $D$, assumed to be known and fixed, represents the distance between the target speaker and the far-field microphone array, $a$ is the speed of sound, $H$ is the hop size of the STFT operation, and $\left \lceil \cdot \right \rceil $ denotes the ceiling function. By calculating Eq. (\ref{eq:num_taps}), the order of filter $\acute{\boldsymbol{h}}_{q}^{d}(f)$ can be directly determined with no need for experimental tuning.

The weighting term $\lambda (t,f)$ is defined as
\small % 这里设置整个公式的字体大小
\begin{equation}
\begin{split}
\lambda (t,f)=\max (\varepsilon \times \max (\left | \boldsymbol{G}_{q} \right |^{2}), \left | G_{q}(t, f) \right |^{2}),
\end{split}
\label{eq:lambda}
\end{equation}
\normalsize % 恢复正常字体大小
where $\left | \boldsymbol{G}_{q} \right |^{2}$ denotes the power spectrogram of $G_{q}(t, f)$, $\max (\cdot )$ calculate the maximum value of a spectrogram, $\varepsilon$ (set to 0.01 in our experiments) is a weight flooring factor and $\max (\cdot ,\cdot )$ returns the bigger one of two values.

To prove the rationality of using Eq. (\ref{eq:filter_est}) to estimate the filter $\hat{\boldsymbol{h}}_{q}^{d}(f)$, we derive Eq. (\ref{eq:filter_est}) as follows: given that $G_{q}(t, f) = R_{q}(t, f)+I_{q}(t, f)$,
\small
\begin{equation}
\begin{split}
&\underset{\acute{\boldsymbol{h}}_{q}^{d}(f)}{\operatorname{argmin}} \sum_{t} \frac{\left|R_{q}(t, f)+I_{q}(t, f)-\acute{\boldsymbol{h}}_{q}^{d}(f)^{\mathrm{H}} \breve{\boldsymbol{Y}}_{q}(t, f)\right|^{2}}{\lambda(t, f)} \\
&\approx \underset{\acute{\boldsymbol{h}}_{q}^{d}(f)}{\operatorname{argmin}} \sum_{t} \frac{\left|R_{q}(t, f)+I_{q}(t, f)-\acute{\boldsymbol{h}}_{q}^{d}(f)^{\mathrm{H}} \breve{\boldsymbol{S}}_{q}(t, f)\right|^{2}}{\lambda(t, f)} \\
&=\underset{\acute{\boldsymbol{h}}_{q}^{d}(f)}{\operatorname{argmin}} \sum_{t} \frac{\Bigl| R_{q}(t, f)-\acute{\boldsymbol{h}}_{q}^{d}(f)^{\mathrm{H}} \breve{\boldsymbol{S}}_{q}(t, f) \Bigr|^{2} + \Bigl| I_{q}(t, f) \Bigr|^{2} }{\lambda(t, f)} \\
% &=\underset{\acute{\boldsymbol{h}}_{q}^{d}(f)}{\operatorname{argmin}} \sum_{t} \text{\scriptsize$\frac{\left|\boldsymbol{h}_{q}^{d}(f)^{\mathrm{H}} \bar{\boldsymbol{S}}_{q}(t, f)-\acute{\boldsymbol{h}}_{q}^{d}(f)^{\mathrm{H}} \breve{\boldsymbol{S}}_{q}(t, f)+\boldsymbol{h}_{q}^{e,l}(f)^{\mathrm{H}} \widetilde{\boldsymbol{S}}_{q}(t, f) \right|^{2}}{\lambda(t, f)}$} \\
&=\underset{\acute{\boldsymbol{h}}_{q}^{d}(f)}{\operatorname{argmin}} \sum_{t} \frac{\text{\scriptsize$\left|\boldsymbol{h}_{q}^{d}(f)^{\mathrm{H}} \bar{\boldsymbol{S}}_{q}(t, f)-\acute{\boldsymbol{h}}_{q}^{d}(f)^{\mathrm{H}} \breve{\boldsymbol{S}}_{q}(t, f)+\boldsymbol{h}_{q}^{e,l}(f)^{\mathrm{H}} \widetilde{\boldsymbol{S}}_{q}(t, f) \right|^{2}$}}{\lambda(t, f)}. \\
\end{split}
\label{eq:err_analyze}
\end{equation}
\normalsize % 恢复正常字体大小
First, based on Eq. (\ref{eq:near_signal}), we obtain the approximate result for the second row. Then, following the W-disjoint orthogonality assumption \cite{yilmaz2004blind}, i.e. the T-F representations of the sources do not overlap, we get the third row. Finally, we decompose $R_{q}(t, f)$ into $\boldsymbol{h}_{q}^{d}(f)^{\mathrm{H}} \bar{\boldsymbol{S}}_{q}(t, f)$ and $\boldsymbol{h}_{q}^{e,l}(f)^{\mathrm{H}} \widetilde{\boldsymbol{S}}_{q}(t, f)$, obtaining the fourth row. Provided that the estimate of $L$ (i.e., $\hat{L}$) is accurate, the term $\boldsymbol{h}_{q}^{e,l}(f)^{\mathrm{H}} \widetilde{\boldsymbol{S}}_{q}(t, f)$ will not affect the estimation of the filter $\hat{\boldsymbol{h}}_{q}^{d}(f)$. Ideally, the estimated filter $\hat{\boldsymbol{h}}_{q}^{d}(f)$ is essentially the relative transfer function (RTF) relating the dry source signal to the direct sound signal.

Ultimately, the oracle direct sound estimated by DSE can be formulated as $X_{q}(t, f) = \hat{\boldsymbol{h}}_{q}^{d}(f)^{\mathrm{H}} \breve{\boldsymbol{Y}}_{q}(t, f)$.

\setlength{\abovecaptionskip}{3pt}  % 调整为你想要的距离
\begin{table}[t]
    \caption{Ablation results of SuPseudo on MISP2023 dataset}
    \renewcommand{\arraystretch}{1.4} % 调整行高
    \setlength{\tabcolsep}{4pt} % 减小表格列之间的默认间距
    \begin{adjustbox}{width=\columnwidth} % 调整表格宽度适应单栏
    \begin{threeparttable}
    \begin{tabular}{ccccccccc}
    \toprule
    \multirow{2}{*}{\textbf{ID}} & \multirow{2}{*}{\textbf{\begin{tabular}[c]{@{}c@{}}Training\\Method\end{tabular}}} & \multirow{2}{*}{\textbf{\begin{tabular}[c]{@{}c@{}}Using\\ SimuData\end{tabular}}} & \multirow{2}{*}{\textbf{Pseudo-Labels}} & \multirow{2}{*}{\textbf{Loss}} & \multirow{2}{*}{\textbf{\begin{tabular}[c]{@{}c@{}}CER\\ (\%)↓\end{tabular}}} & \multicolumn{3}{c}{\textbf{DNSMOS P.835} \cite{reddy2022dnsmos}} \\ \cline{7-9} 
    &  &  &  &  &  & \textbf{SIG↑} & \textbf{BAK↑} & \textbf{OVRL↑} \\ 
    \toprule
    B1 & - & - & - & - & 37.60 & 1.68 & 1.63 & 1.31 \\
    \hline
    M2 & supervised & \ding{51} & - & - & 36.64 & 1.88 & 1.92 & 1.40 \\
    \hline
    M3 & SuPseudo & \ding{55} & DSE signals & $\mathcal{L} _\mathrm{MCA}$ & 30.44 & 2.22 & 2.74 & 1.69 \\ 
    \hline
    M4a & SuPseudo & \ding{51} & Close-talk Segs. & $\mathcal{L} _\mathrm{MSE}$ & 34.02 & 1.81 & 1.87 & 1.37 \\
    M4b & SuPseudo & \ding{51} & Close-talk Segs. & $\mathcal{L} _\mathrm{MCA}$ & 33.26 & 1.87 & 1.95 & 1.41 \\
    \hdashline
    M4c & SuPseudo & \ding{51} & DSE signals & $\mathcal{L} _\mathrm{MSE}$ & 29.99 & 2.22 & \textbf{2.74} & 1.69 \\
    M4d & SuPseudo & \ding{51} & DSE signals & $\mathcal{L} _\mathrm{MCA}$ & \textbf{29.80} & \textbf{2.22} & 2.73 & \textbf{1.69} \\
    \bottomrule
    \end{tabular}
    \begin{tablenotes}    %这行要添加， 从这开始
        \footnotesize               %这行要添加
        \item \textit{Notes}: System B1 refers to “GSS”, while systems M2-M4d refer to “GSS+FARNET”. The ASR back-end used for decoding is the official ASR model as described in \cite{dai2023improving}.
    \end{tablenotes}            %这行要添加
    \end{threeparttable}       %这行要添加，到这里结束
    \end{adjustbox}
    \label{table1}
    \vspace{-15pt} % 减少图片下方的间距
\end{table}

\subsection{MCA Loss}
Given that the assumptions in Eq. (\ref{eq:err_analyze}) are not fully met, the DSE-estimates (i.e., pseudo-labels) may not be accurate exactly. To improve robustness during training on real-recorded data, we propose \textit{magnitude constraint adjustable} (MCA) loss. The MCA loss is obtained by linearly weighting MSE loss $\mathcal{L} _\mathrm{MSE}$ and cosine similarity loss $\mathcal{L} _\mathrm{COSSIM}$ as follows:
\small % 这里设置整个公式的字体大小
\begin{equation}
\begin{split}
\mathcal{L} _\mathrm{MSE} &= \mathbb{E}_{\boldsymbol{A},\boldsymbol{B}}\left [ || \boldsymbol{A}-\boldsymbol{B} ||_{F}^{2}  \right ],   \\
\mathcal{L} _\mathrm{COSSIM} &= 1-\cos\left ( \boldsymbol{A},\boldsymbol{B} \right ) = 1 - \frac{<\boldsymbol{A},\boldsymbol{B}>_{F} }{|| \boldsymbol{A} ||_{F} || \boldsymbol{B} ||_{F}}, \\
\mathcal{L} _\mathrm{MCA} &= \mathcal{L} _\mathrm{MSE} + \alpha \times \mathcal{L} _\mathrm{COSSIM},
\end{split}
\label{eq:CAL}
\end{equation}
\normalsize % 恢复正常字体大小
where $\boldsymbol{A}$ and $\boldsymbol{B}$ respectively denote the oracle and the estimated magnitude spectrograms (e.g., ${\left | \boldsymbol{X}_{q}  \right | ^{c}}$ and $\widehat{\left | \boldsymbol{X}_{q}  \right | ^{c}}$). $\cos\left ( \cdot ,\cdot  \right )$ computes the cosine similarity between two magnitude spectrograms, $<\cdot ,\cdot >_{F}$ denotes the Frobenius inner product, and $\alpha$ is a regulator that controls the weight of $\mathcal{L} _\mathrm{COSSIM}$. Unlike $\mathcal{L} _\mathrm{MSE}$, $\mathcal{L} _\mathrm{COSSIM}$ does not strictly force $\boldsymbol{A}$ and $\boldsymbol{B}$ to be identical (i.e., $\mathcal{L} _\mathrm{COSSIM}$ is a softer constraint). Therefore, its introduction makes $\mathcal{L} _\mathrm{MCA}$ more robust when dealing with misaligned data pairs and also allows the model to pay more attention to the similarity between $\boldsymbol{A}$ and $\boldsymbol{B}$.

\setlength{\abovecaptionskip}{3pt}  % 调整为你想要的距离
\begin{table*}[htp]
    \caption{Comparison with other front-end systems on MISP2023 evaluation set}
    \renewcommand{\arraystretch}{1.4} % 调整行高
    \setlength{\tabcolsep}{4pt} % 减小表格列之间的默认间距
    \begin{adjustbox}{width=\textwidth} % 自动调整表格
    \begin{threeparttable}
    \begin{tabular}{@{}cccccccccccc@{}}
    \toprule
    \multirow{2}{*}{\textbf{ID}} & \multirow{2}{*}{\textbf{\begin{tabular}[c]{@{}c@{}}Front-end\\System\end{tabular}}} & \multirow{2}{*}{\textbf{\begin{tabular}[c]{@{}c@{}}Training\\Method\end{tabular}}} & \multirow{2}{*}{\textbf{\begin{tabular}[c]{@{}c@{}}Joint\\ Opt.\end{tabular}}} & \multirow{2}{*}{\textbf{Modality}} & \multirow{2}{*}{\textbf{\begin{tabular}[c]{@{}c@{}}System\\ Fusion\end{tabular}}} & \multicolumn{3}{c}{\textbf{CER(\%)$\downarrow$}} & \multicolumn{3}{c}{\textbf{DNSMOS P.835} \cite{reddy2022dnsmos}} \\ 
    \cmidrule(lr){7-9} \cmidrule(l){10-12} 
     &  &  &  &  &  & \textbf{Official} \cite{dai2023improving} & \textbf{Paraformer} \cite{WOS:000900724502048} & \textbf{Whisper} \cite{radford2023robust} & \textbf{SIG$\uparrow$} & \textbf{BAK$\uparrow$} & \textbf{OVRL$\uparrow$} \\ 
    \toprule
    B1 & GSS \cite{raj23_interspeech} & - & - & A & - & 37.6 & 40.52 & 56.85 & 1.68 & 1.63 & 1.31 \\
    \hline
    B2 & GSS+MEASE \cite{wu2024multimodal} & supervised & \ding{51} & A+V & \ding{55} & 36.09 & 40.26 & 58.87 & 1.89 & 1.99 & 1.41 \\
    % 1b & SPAV-TFGridNet \cite{hou2024sir} & Supervised & - & A+V & \ding{55} & \ding{55} & 34.49 & - & - & - & - & - \\
    \hline
    S1 & XMUSPEECH \cite{luo2024xmuspeech} & supervised & \ding{55} & A+V & \ding{51} & 33.41(\underline{$3^{rd}$}) & 31.70 & 49.40 & 1.91 & 2.07 & 1.51 \\
    S2 & NPU-MSXF \cite{han2024audio} & supervised & \ding{51} & A+V & \ding{51} & 33.21(\underline{$2^{nd}$}) & - & - & - & - & 1.41 \\
    S3 & NJU-AALab \cite{hou2024sir} & supervised & \ding{55} & A+V & \ding{51} & 33.18(\underline{$1^{st}$}) & - & - & - & - & - \\
    \hline
    M4d & GSS+FARNET (ours) & SuPseudo & \ding{55} & A & \ding{55} & \textbf{29.80} & \textbf{27.89} & \textbf{43.77} & \textbf{2.22} & \textbf{2.73} & \textbf{1.69} \\ 
    \bottomrule
    \end{tabular}
    \begin{tablenotes}    %这行要添加， 从这开始
        \footnotesize               %这行要添加
        \item \textit{Notes}: Systems S1-S3 all utilize GSS, along with the oracle diarization, to pre-process raw 6-channel far-field audio. “$(\underline{\hspace{0.05cm}\cdot\hspace{0.05cm}})$” gives the ranking of the challenge.
    \end{tablenotes}            %这行要添加
    \end{threeparttable}       %这行要添加，到这里结束
    \end{adjustbox}
    \label{table2}
    \vspace{-15pt} % 减少图片下方的间距
\end{table*}

% \vspace{-5pt}
\section{Experimental Setup}
For dataset, we utilize the MISP2023 corpus \cite{wu2024multimodal}, a far-field multi-speaker conversational Chinese audio-visual corpus, to evaluate our method. This corpus focuses on real home-TV scenarios (as shown in Fig. \ref{fig1}), where 2-6 speakers (i.e., $N\in\left \{ 2,...,6 \right \}$) spontaneously communicate without specific topics under strong background noise and reverberation. The training set includes 106.09 hours of close-talk/middle-field/far-field audio and middle-field/far-field videos, collected from real 21 rooms and unique 200 speakers. The development and evaluation sets include 2.51 and 3.62 hours of far-field audio and middle-field video, respectively. It should be noted that our method only utilizes the audio data.

For both DSE method and FARNET model, the window size and hop size of STFT are set to 25 ms and 6.25 ms, respectively. The filter order $\hat{L}$ is set to 4 according to Eq. (\ref{eq:num_taps}), with $D=5 \, \text{m}$ \cite{wu2024multimodal}, $a=340 \, \text{m/s}$, and $H=6.25 \, \text{ms}$. The weighting factor $\alpha$ of $\mathcal{L} _\mathrm{MCA}$ is set to 0.2. For data preparation, 243.19 hours of simulated data is generated using official method \cite{wu2024multimodal}. For data pre-processing, we use GSS, along with the oracle diarization, to initially separate raw far-field mixtures of both real-recorded data (i.e., the official training set) and simulated data. And then we use the proposed DSE to generate label-estimates of real-recorded data. For two-stage SuPseudo training, an initial learning rate of 1.75e-3, a batch size of 7, and a cut-length of 2 seconds are used in both stages. And our FARNET model is trained for 40 epochs in stage 1 and 8 epochs in stage 2.  

For evaluation metrics, we follow the official guidelines. The enhanced speech is fed to the official ASR model \cite{dai2023improving} for decoding, and then the character error rate (CER) is calculated for ranking. In addition to the official ASR model \cite{dai2023improving} trained on the MISP2023 corpus, we additionally introduce two universal ASR models for decoding: Paraformer-large \cite{WOS:000900724502048} trained on 60,000 hours of mandarin data, and Whisper-large-v3 \cite{radford2023robust} trained on 680,000 hours of multilingual data. Additionally, since the ground-truth target speech reference is unavailable for the MISP2023 dataset, the DNSMOS P.835 \cite{reddy2022dnsmos}, a non-intrusive perceptual objective metric, is used to evaluate enhanced speech quality.   

For comparison, GSS \cite{raj23_interspeech} is one of the most commonly used front-ends in far-field speech recognition and also serves as the baseline before further enhancement. We also consider the top-ranking systems from the MISP2023 Challenge\footnote{\url{https://mispchallenge.github.io/mispchallenge2023/index.html}}, including NJU-system \cite{hou2024sir}, NPU-MSXF \cite{han2024audio}, and XMU-system \cite{luo2024xmuspeech}. It should be noted that all these top-ranking systems employ multi-system fusion strategy. To exclude the influence of multi-system fusion, we reference the single-system method, the official baseline “GSS+MEASE” \cite{wu2024multimodal}.

% \vspace{-5pt}
\section{Results And Analysis}
In Table \ref{table1}, the impact of various training strategies on model performance is illustrated. It is clear that systems M2-M4d all outperform GSS. This suggests that the GSS-output still contains significant interferences and highlights the effectiveness of the FARNET model. While system M2, which employs a supervised learning method with massive simulated data (SimuData), has already surpassed GSS, our proposed SuPseudo method with only real-recorded data (system M3) achieves a much lower CER (with a relative reduction of 17\%) and much higher DNSMOS scores (with relative improvements of 18\%, 43\%, and 21\% on SIG, BAK, and OVRL, respectively), clearly demonstrating the effectiveness of pseudo-supervised learning. And apparently, system M4d achieves the best performance, demonstrating that two-stage training strategy effectively integrate the benefits of both simulated and real-recorded data. In systems M4a-M4b, the close-talk speech segments (Close-talk Segs.) are directly used as pseudo-labels for fine-tuning, still showing a performance improvement over system M2, supporting the rationality of using close-talk speech segments to estimate the oracle direct sound. Additionally, from system M4a vs. M4b and M4c vs. M4d, it is obvious that the $\mathcal{L} _\mathrm{MCA}$ brings further relative improvement when there are larger errors in pseudo-labels (in the case of system M4a vs. M4b). 

In Table \ref{table2}, we compare our “GSS+FARNET” system trained using SuPseudo (i.e., system M4d in Table \ref{table1}) with the official baseline and the top-ranking systems from the MISP2023 Challenge. Comparing system B2 with M4d, it can be found that although FARNET model utilizes only audio modality (audio vs. audio-visual) and has significantly fewer parameters (1.56M vs. 118.95M), it still greatly outperforms MEASE on all evaluation metrics. This further demonstrates the effectiveness of both FARNET model and SuPseudo training. Additionally, FARNET maintains optimal performance across all three ASR back-ends, which we attribute to the signal-level supervision provided by SuPseudo. In contrast, MEASE, jointly optimized with the official ASR model, exhibits performance degradation on two non-official ASR models, Paraformer-large and Whisper-large-v3. This indicates that the joint optimization does not fundamentally resolve the domain mismatch issue. Furthermore, compared with the top-ranking multi-system method (S1-S3), our single-system “GSS+FARNET” trained using SuPseudo achieves superior performance, with a 10.2\% relative reduction on CER over the previous SOTA.

% \vspace{-5pt}
\section{Conclusion}
In this paper, we propose DSE to estimate the oracle direct sound of real-recorded data for SE. And we present SuPseudo, a novel training framework to leaveage the DSE-estimates and enhance the models’ generalization capability to real-world conditions. To better match far-field speech recognition, we adopt the FARNET model as the backbone. Ablation experiments and comparisons with the top-ranking systems on the MISP2023 corpus demonstrate the effectiveness of our method. Our future work will focus on replacing “GSS+FARNET” system with a single target speaker extraction (TSE) model, in conjunction with SuPseudo, to develop a fully data-driven front-end algorithm for real-world far-field speech recognition.

% \section{Acknowledgments}
% This work was supported in part by the National Natural Science Foundation of China under Grants 62371407 and 62276220, and the Innovation of Policing Science and Technology, Fujian province (Grant number: 2024Y0068).

% \small
\bibliographystyle{IEEEtran}
\bibliography{mybib}

% Generated by IEEEtran.bst, version: 1.13 (2008/09/30)
\begin{thebibliography}{10}
\providecommand{\url}[1]{#1}
\csname url@samestyle\endcsname
\providecommand{\newblock}{\relax}
\providecommand{\bibinfo}[2]{#2}
\providecommand{\BIBentrySTDinterwordspacing}{\spaceskip=0pt\relax}
\providecommand{\BIBentryALTinterwordstretchfactor}{4}
\providecommand{\BIBentryALTinterwordspacing}{\spaceskip=\fontdimen2\font plus
\BIBentryALTinterwordstretchfactor\fontdimen3\font minus \fontdimen4\font\relax}
\providecommand{\BIBforeignlanguage}[2]{{%
\expandafter\ifx\csname l@#1\endcsname\relax
\typeout{** WARNING: IEEEtran.bst: No hyphenation pattern has been}%
\typeout{** loaded for the language `#1'. Using the pattern for}%
\typeout{** the default language instead.}%
\else
\language=\csname l@#1\endcsname
\fi
#2}}
\providecommand{\BIBdecl}{\relax}
\BIBdecl

\bibitem{boeddeker2018front}
C.~Boeddeker, J.~Heitkaemper, J.~Schmalenstroeer, L.~Drude, J.~Heymann, and R.~Haeb-Umbach, ``Front-end processing for the chime-5 dinner party scenario,'' in \emph{CHiME5 Workshop, Hyderabad, India}, vol.~1, 2018.

\bibitem{hao2021fullsubnet}
X.~Hao, X.~Su, R.~Horaud, and X.~Li, ``Fullsubnet: A full-band and sub-band fusion model for real-time single-channel speech enhancement,'' in \emph{ICASSP 2021-2021 IEEE International Conference on Acoustics, Speech and Signal Processing (ICASSP)}.\hskip 1em plus 0.5em minus 0.4em\relax IEEE, 2021, pp. 6633--6637.

\bibitem{yu2023efficient}
J.~Yu and Y.~Luo, ``Efficient monaural speech enhancement with universal sample rate band-split rnn,'' in \emph{ICASSP 2023-2023 IEEE International Conference on Acoustics, Speech and Signal Processing (ICASSP)}.\hskip 1em plus 0.5em minus 0.4em\relax IEEE, 2023, pp. 1--5.

\bibitem{abdulatif2024cmgan}
S.~Abdulatif, R.~Cao, and B.~Yang, ``Cmgan: Conformer-based metric-gan for monaural speech enhancement,'' \emph{IEEE/ACM Transactions on Audio, Speech, and Language Processing}, 2024.

\bibitem{wu2024multimodal}
S.~Wu, C.~Wang, H.~Chen, Y.~Dai, C.~Zhang, R.~Wang, H.~Lan, J.~Du, C.-H. Lee, J.~Chen \emph{et~al.}, ``The multimodal information based speech processing (misp) 2023 challenge: Audio-visual target speaker extraction,'' in \emph{ICASSP 2024-2024 IEEE International Conference on Acoustics, Speech and Signal Processing (ICASSP)}.\hskip 1em plus 0.5em minus 0.4em\relax IEEE, 2024, pp. 8351--8355.

\bibitem{vinnikov2024notsofar}
A.~Vinnikov, A.~Ivry, A.~Hurvitz, I.~Abramovski, S.~Koubi, I.~Gurvich, S.~Peer, X.~Xiao, B.~M. Elizalde, N.~Kanda \emph{et~al.}, ``Notsofar-1 challenge: New datasets, baseline, and tasks for distant meeting transcription,'' \emph{arXiv preprint arXiv:2401.08887}, 2024.

\bibitem{barker2018fifth}
J.~Barker, S.~Watanabe, E.~Vincent, and J.~Trmal, ``The fifth'chime'speech separation and recognition challenge: Dataset, task and baselines,'' \emph{Interspeech 2018}, 2018.

\bibitem{watanabe2020chime}
S.~Watanabe, M.~Mandel, J.~Barker, E.~Vincent, A.~Arora, X.~Chang, S.~Khudanpur, V.~Manohar, D.~Povey, D.~Raj \emph{et~al.}, ``Chime-6 challenge: Tackling multispeaker speech recognition for unsegmented recordings,'' in \emph{CHiME 2020-6th International Workshop on Speech Processing in Everyday Environments}, 2020.

\bibitem{cornell2023chime}
S.~Cornell, M.~Wiesner, S.~Watanabe, D.~Raj, X.~Chang, P.~Garcia, M.~Maciejewski, Y.~Masuyama, Z.-Q. Wang, S.~Squartini \emph{et~al.}, ``The chime-7 dasr challenge: Distant meeting transcription with multiple devices in diverse scenarios,'' \emph{arXiv preprint arXiv:2306.13734}, 2023.

\bibitem{chen2022first}
H.~Chen, H.~Zhou, J.~Du, C.-H. Lee, J.~Chen, S.~Watanabe, S.~M. Siniscalchi, O.~Scharenborg, D.-Y. Liu, B.-C. Yin \emph{et~al.}, ``The first multimodal information based speech processing (misp) challenge: Data, tasks, baselines and results,'' in \emph{ICASSP 2022-2022 IEEE International Conference on Acoustics, Speech and Signal Processing (ICASSP)}.\hskip 1em plus 0.5em minus 0.4em\relax IEEE, 2022, pp. 9266--9270.

\bibitem{wang2023multimodal}
Z.~Wang, S.~Wu, H.~Chen, M.-K. He, J.~Du, C.-H. Lee, J.~Chen, S.~Watanabe, S.~Siniscalchi, O.~Scharenborg \emph{et~al.}, ``The multimodal information based speech processing (misp) 2022 challenge: Audio-visual diarization and recognition,'' in \emph{ICASSP 2023-2023 IEEE International Conference on Acoustics, Speech and Signal Processing (ICASSP)}.\hskip 1em plus 0.5em minus 0.4em\relax IEEE, 2023, pp. 1--5.

\bibitem{michelsanti2019effects}
D.~Michelsanti, Z.-H. Tan, S.~Sigurdsson, and J.~Jensen, ``Effects of lombard reflex on the performance of deep-learning-based audio-visual speech enhancement systems,'' in \emph{ICASSP 2019-2019 IEEE International Conference on Acoustics, Speech and Signal Processing (ICASSP)}.\hskip 1em plus 0.5em minus 0.4em\relax IEEE, 2019, pp. 6615--6619.

\bibitem{leglaive2023chime}
S.~Leglaive, L.~Borne, E.~Tzinis, M.~Sadeghi, M.~Fraticelli, S.~Wisdom, M.~Pariente, D.~Pressnitzer, and J.~R. Hershey, ``The chime-7 udase task: Unsupervised domain adaptation for conversational speech enhancement,'' in \emph{7th International Workshop on Speech Processing in Everyday Environments (CHiME)}, 2023.

\bibitem{han2024audio}
R.~Han, X.~Yan, W.~Xu, P.~Guo, J.~Sun, H.~Wang, Q.~Lu, N.~Jiang, and L.~Xie, ``An audio-quality-based multi-strategy approach for target speaker extraction in the misp 2023 challenge,'' in \emph{2024 IEEE International Conference on Acoustics, Speech, and Signal Processing Workshops (ICASSPW)}.\hskip 1em plus 0.5em minus 0.4em\relax IEEE, 2024, pp. 27--28.

\bibitem{xu2024employing}
Z.~Xu, M.~Sach, J.~Pirklbauer, and T.~Fingscheidt, ``Employing real training data for deep noise suppression,'' in \emph{ICASSP 2024-2024 IEEE International Conference on Acoustics, Speech and Signal Processing (ICASSP)}.\hskip 1em plus 0.5em minus 0.4em\relax IEEE, 2024, pp. 10\,731--10\,735.

\bibitem{reddy2022dnsmos}
C.~K. Reddy, V.~Gopal, and R.~Cutler, ``Dnsmos p. 835: A non-intrusive perceptual objective speech quality metric to evaluate noise suppressors,'' in \emph{ICASSP 2022-2022 IEEE International Conference on Acoustics, Speech and Signal Processing (ICASSP)}.\hskip 1em plus 0.5em minus 0.4em\relax IEEE, 2022, pp. 886--890.

\bibitem{wang2024superm2m}
\BIBentryALTinterwordspacing
Z.-Q. Wang, ``Superm2m: Supervised and mixture-to-mixture co-learning for speech enhancement and robust asr,'' 2024. [Online]. Available: \url{https://arxiv.org/abs/2403.10271}
\BIBentrySTDinterwordspacing

\bibitem{lv2021pseudo}
J.~Lv, Z.~Kang, X.~Lu, and Z.~Xu, ``Pseudo-supervised deep subspace clustering,'' \emph{IEEE Transactions on Image Processing}, vol.~30, pp. 5252--5263, 2021.

\bibitem{raj23_interspeech}
D.~Raj, D.~Povey, and S.~Khudanpur, ``Gpu-accelerated guided source separation for meeting transcription,'' in \emph{Interspeech 2023}, 2023, pp. 3507--3511.

\bibitem{talmon2009relative}
R.~Talmon, I.~Cohen, and S.~Gannot, ``Relative transfer function identification using convolutive transfer function approximation,'' \emph{IEEE Transactions on audio, speech, and language processing}, vol.~17, no.~4, pp. 546--555, 2009.

\bibitem{braun2021consolidated}
S.~Braun and I.~Tashev, ``A consolidated view of loss functions for supervised deep learning-based speech enhancement,'' in \emph{2021 44th International Conference on Telecommunications and Signal Processing (TSP)}.\hskip 1em plus 0.5em minus 0.4em\relax IEEE, 2021, pp. 72--76.

\bibitem{luo2024xmuspeech}
L.~Luo, T.~Li, L.~Li, and Q.~Hong, ``The xmuspeech system for audio-visual target speaker extraction in misp 2023 challenge,'' in \emph{2024 IEEE International Conference on Acoustics, Speech, and Signal Processing Workshops (ICASSPW)}.\hskip 1em plus 0.5em minus 0.4em\relax IEEE, 2024, pp. 39--40.

\bibitem{zhang2016deep}
X.-L. Zhang and D.~Wang, ``A deep ensemble learning method for monaural speech separation,'' \emph{IEEE/ACM transactions on audio, speech, and language processing}, vol.~24, no.~5, pp. 967--977, 2016.

\bibitem{pandey2020densely}
A.~Pandey and D.~Wang, ``Densely connected neural network with dilated convolutions for real-time speech enhancement in the time domain,'' in \emph{ICASSP 2020-2020 IEEE International Conference on Acoustics, Speech and Signal Processing (ICASSP)}.\hskip 1em plus 0.5em minus 0.4em\relax IEEE, 2020, pp. 6629--6633.

\bibitem{ulyanov2016instance}
D.~Ulyanov, A.~Vedaldi, and V.~Lempitsky, ``Instance normalization: The missing ingredient for fast stylization,'' \emph{arXiv preprint arXiv:1607.08022}, 2016.

\bibitem{wang2021convolutive}
Z.-Q. Wang, G.~Wichern, and J.~Le~Roux, ``Convolutive prediction for monaural speech dereverberation and noisy-reverberant speaker separation,'' \emph{IEEE/ACM Transactions on Audio, Speech, and Language Processing}, vol.~29, pp. 3476--3490, 2021.

\bibitem{yilmaz2004blind}
O.~Yilmaz and S.~Rickard, ``Blind separation of speech mixtures via time-frequency masking,'' \emph{IEEE Transactions on signal processing}, vol.~52, no.~7, pp. 1830--1847, 2004.

\bibitem{dai2023improving}
Y.~Dai, H.~Chen, J.~Du, X.~Ding, N.~Ding, F.~Jiang, and C.-H. Lee, ``Improving audio-visual speech recognition by lip-subword correlation based visual pre-training and cross-modal fusion encoder,'' in \emph{2023 IEEE International Conference on Multimedia and Expo (ICME)}.\hskip 1em plus 0.5em minus 0.4em\relax IEEE, 2023, pp. 2627--2632.

\bibitem{WOS:000900724502048}
Z.~Gao, S.~Zhang, I.~McLoughlin, and Z.~Yan, ``Paraformer: Fast and accurate parallel transformer for non-autoregressive end-to-end speech recognition,'' in \emph{Interspeech 2022}, ser. Interspeech, 2022, pp. 2063--2067, interspeech Conference, Incheon, SOUTH KOREA, SEP 18-22, 2022.

\bibitem{radford2023robust}
A.~Radford, J.~W. Kim, T.~Xu, G.~Brockman, C.~McLeavey, and I.~Sutskever, ``Robust speech recognition via large-scale weak supervision,'' in \emph{International conference on machine learning}.\hskip 1em plus 0.5em minus 0.4em\relax PMLR, 2023, pp. 28\,492--28\,518.

\bibitem{hou2024sir}
Z.~Hou, T.~Sun, Y.~Hu, C.~Zhu, K.~Chen, and J.~Lu, ``Sir-progressive audio-visual tf-gridnet with asr-aware selector for target speaker extraction in misp 2023 challenge,'' in \emph{2024 IEEE International Conference on Acoustics, Speech, and Signal Processing Workshops (ICASSPW)}.\hskip 1em plus 0.5em minus 0.4em\relax IEEE, 2024, pp. 11--12.

\end{thebibliography}

\end{document}